\documentclass[12pt]{article}
\usepackage{graphicx}
\usepackage{pazha}

\usepackage{lscape}
\usepackage{amssymb}
\usepackage{amsmath}
\usepackage[T2A]{fontenc}

\tightenlines

\def\arcsec{$^{\prime\prime\,}$}
\def\arcmin{$^{\prime\,}$}
\def\fdg{\hbox{$.\!\!^\circ$}}

\def\AA{\buildrel _{\hskip 0.5pt \circ} \over {\mathrm{A}}}

\def\smfigure#1#2#3{
  \begin{minipage}{0.9\columnwidth}
    \begin{minipage}{0.049\columnwidth}
      \rotatebox{90}{\phantom{0000}#3}
    \end{minipage}
    \begin{minipage}{0.9\columnwidth}
      \includegraphics[bb=40 188 556 678,width=0.97\columnwidth]{#1}
      \centerline{#2}
    \end{minipage}

    \vskip 3pt
    ~
  \end{minipage}
}
\begin{document}

{\it Will be published in ''Astronomy Letters'', 2010, V.36, N12, pp.904-909}

\bigskip

\title{\bf IGR\,J16547$-$1916/1RXS\,J165443.5$-$191620 --- a New Intermediate
Polar from the INTEGRAL Galactic Survey}

\author{\bf Lutovinov A.\affilmark{1}$^{\,*}$, Burenin R.\affilmark{1}, Revnivtsev M.\affilmark{1}, Suleimanov V.\affilmark{2,3}, Tkachenko A.\affilmark{1}}

\affil{$^1$ {\it Space Research Institute, Moscow,
117997, Profsoyuznaya str. 84/32, Russia}\\}

\affil{$^2$ {\it T\"ubingen University, T\"ubingen, Germany}\\}

\affil{$^3$ {\it Kazan (Volga Region) Federal University, Kremlevskaya
ul. 18, Kazan, 420008 Russia}\\}

\vspace{5mm}

Received July 1, 2010

\sloppypar
\vspace{2mm}
\noindent

We present the results of our optical identification of the X-ray source
IGR\, J16547-1916 detected by the INTEGRAL observatory during a deep all-sky
survey. Analysis of the spectroscopic data from the SWIFT and INTEGRAL
observatories in the X-ray energy band and from the BTA (Special
Astrophysical Observatory) telescope in the optical band has shown that the
source is most likely an intermediate polar -- an accreting white dwarf with
the mass of $M_{\rm WD}\simeq0.85 M_{\odot}$ in a low-mass binary
system. Subsequent studies of the object's rapid variability with the
RTT-150 telescope have confirmed this conclusion by revealing periodic
pulsations of its optical emission with a period of $\approx550$ s.


\vfill

{$^{*}$ e-mail: aal@iki.rssi.ru}

\section*{INTRODUCTION}

The surveys that make it possible to measure the space density of binary
systems with accreting white dwarfs are very rare. This is mainly because
such objects are detected mostly in the optical band according to various
criteria that can barely be described quantitatively (outbursts,
variability, eclipses, etc). X-ray sky surveys have significant advantages
in this sense. Such surveys have well-defined limits of sensitivity to
sources in different parts of the sky, which allows the densities of various
components of the Galactic stellar population to be estimated more reliably
(see, e.g., Hertz and Grindlay 1984; Verbunt et al. 1997; Grimm et al. 2002;
Revnivtsev et al. 2008).  One of the currently best Galactic hard X-ray
surveys is the INTEGRAL survey (see, e.g., Krivonos et al 2007, 2010; Bird
et al. 2010). However, to carry out statistical studies and to make reliable
estimates of the densities of objects using the survey results, it is
necessary that the completeness of the optical identifications of cataloged
objects be high enough.  For this purpose, several groups in the world
systematically perform observations of the objects detected by the INTEGRAL
observatory in an effort to establish their nature (see, e.g., Bikmaev et
al. 2006, 2008; Burenin et al. 2008, 2009; Sazonov et al. 2008; Masetti et
al. 2007, 2010; Tomsik et al. 2008, 2009).  Persistent bright hard X-ray
sources in our Galaxy are not that many -- about 180, according to the
latest catalog by Krivonos et al. (2010); only 33 of them are accreting
white dwarfs. This means that determining the nature of each Galactic source
of such a brightness is of considerable value.

The source IGR\,J16547-1916 has been detected at a statistically significant
level by INTEGRAL only in the sum map obtained over seven years of its in-
orbit operation with a flux of $0.74\pm0.12$ mCrab ($\sim 10^{-11}$ erg
s$^{-1}$ cm$^{-2}$) in the $17-60$ keV energy band (Krivonos et al. 2010;
see also Bird et al. 2010). The source has been previously observed in the
ROSAT all-sky survey and is designated as 1RXS\,J165443.5-191620.  Here,
based on a comprehensive analysis of the spectroscopic and timing data from
the INTEGRAL and SWIFT X-ray observatories and the optical BTA (Special
Astrophysical Observatory, SAO) and RTT-150 Russian-Turkish (TUBITAK
Observatory, Turkey) telescopes, we show that IGR\,J16547-1916 is most
probably an intermediate polar -- an accreting white dwarf whose magnetic
field is strong enough to destroy the disk flow near the white dwarf surface
but is not strong enough to completely prevent its formation (see, e.g.,
Patterson 1994).

\section{OPTICAL IDENTIFICATION}

The source IGR\,J16547-1916 is detected at a statistically significant level
in the sum all-sky map obtained by INTEGRAL over seven years of its in-orbit
operation. The effective exposure time of the sky region under consideration
was $\sim1.9$ Ms. The position of the source on the celestial sphere was
determined from these data with an uncertainty of 3\arcmin (Krivonos et
al. 2010).

To determine the source position with a higher accuracy (up to arcseconds)
and to construct its soft ($< 10$ keV, see below) X-ray spectrum, we used
the XRT/SWIFT data obtained on January 21 and 23, 2010, (ObsID 90182) in the
Photon Counting mode with a total exposure time of $\sim5$ ks. The source
was detected at a statistically significant level in both observations with
approximately the same flux, which allowed us to refine its coordinates,
RA=16$^h$ 54$^m$ 43.7$^s$, Dec=-19\fdg 16\arcmin 30\arcsec (J2000), and to
improve the accuracy of their determination $\sim3''$. One optical object
with a magnitude $R\approx15.3$ (see Fig. 1) visible on the digitized
Palomar Observatory Sky Survey plates falls with confidence into the error
box of IGR\,J16547-1916.  To establish the nature of the source, its
spectroscopic observations were performed with the 6-m BTA telescope (SAO,
the Russian Academy of Sciences) in June 2010. The spectroscopic
observations were performed on the night of June 9, 2010, with the SCORPIO
spectrometer (Afanasiev and Moiseev 2005). A 1\arcsec\--wide long slit and a
400 lines mm$^{-1}$ diffraction grating were used in the observations; the
total exposure time was 600 s. This allowed a highquality spectrum of the
source to be obtained in the range from 3900 to $8500\AA$ with a resolution
of $\approx16\AA$.

The optical spectrum of this object is shown in Fig. 2. It exhibits a blue
continuum and a set of intense hydrogen and helium emission lines that are
typical of the spectra of accretion disks around white dwarfs (see, e.g.,
Williams and Fergusson 1982). Thus, this object is a cataclysmic variable --
an accreting white dwarf in a binary system. Recently, Masetti et al. (2010)
have reached the same conclusion based on optical spectroscopic
observations.

\section{OPTICAL VARIABILITY}

To investigate the temporal variability of the optical emission from
IGR\,J16547-1916, photometric observations of this object were performed
with the RTT-150 telescope. The observations were performed with the medium-
and low-resolution TFOCS spectrometer on the night of June 27, 2010, for 2.3
h. A total of 297 measurements of the optical flux from the source were
obtained with a time resolution of about 30 s. The Lomb-Scargle periodogram
of the derived light curve (see Fig. 3a) revealed brightness oscillations in
the source with a period of about $549\pm15$ s (see Fig. 3b). The brightness
oscillations were coherent throughout the observations (as far as this can
be judged from a $\approx$8000-s-long time series).

The detected brightness oscillations can be caused either by the orbital
brightness variability in the binary system or by the variability related to
the white dwarf spin. An orbital period of $\sim549$ s seems too short for a hard
X-ray source; binary systems with such short orbital periods usually accrete
matter without disk formation and have a fairly soft X-ray spectrum (see
V407 Vul, ES Cet, AM CVn). In our case, we are most likely dealing with the
white dwarf spin period and the orbital period of the binary system is
considerably longer than 549 s. Consequently, IGR\,J16547-1916 is an
intermediate polar.

\section{X-RAY SPECTROSCOPY}

Figure 4 shows a broadband spectrum of
IGR\,J16547-1916/1RXS\,J165443.5-191620 in the 0.6$-$120 keV energy band
constructed from SWIFT and INTEGRAL data. It should be noted that, although
the ''soft'' ($<10$ keV) part of the spectrum was obtained in January 2010
and the ''hard'' ($>17$ keV) one was accumulated in the period from 2003 to
2009, they agree well with each other, which may be indicative of a
relatively constant spectral shape.  Therefore, we subsequently fitted the
source spectrum in a wide energy range by a single model, leaving the
normalization of the INTEGRAL spectrum as a free parameter.

On the whole, the spectrum of IGR\,J16547-C1916 can be well described by the
model of partially absorbed (i.e., when only part of the emergent radiation
is intercepted and absorbed by the medium) bremsstrahlung with a temperature
of $\sim21$ keV. The best-fit parameters are given below:

\medskip

\begin{tabular}{l|c|c|c|c|l|l}
$T_{br}$, keV & $K$, & $N_{H}$, 10$^{22}$ cm$^{-2}$ & $C_{F}$ & $F_{2-100}$ & $F_{0.1-100}$ & $\chi^2$ \\[1mm]

\hline
$21.4^{+5.0}_{-7.3}$ & $4.35^{+0.72}_{-0.72}$ & $8.0^{+1.3}_{-1.5}$ & $0.942^{+0.011}_{-0.013}$ & $1.73^{+0.20}_{-0.13}$ & $4.73^{+0.32}_{-1.37}$ & 0.83 \\
\end{tabular}

\medskip

Here, $T_{br}$ is the emitting-plasma (bremsstrahlung) temperature, K is the
normalization of the bremsstrahlung spectrum given in units $10^{-3}$ and
defined by the expression

\begin{equation}
  K = \frac{3.02\times 10^{-15}}{4\pi D^2} \int n^2_e \, dV,
\end{equation}

\noindent
where $D$ is the distance to the source in cm, $n_e$ is the electron number
density in the emitting plasma, and $V$ is its volume; $C_F$ is the fraction
of the absorbed radiation in neutral matter with an equivalent column
density $N_H$, $F_{2-100}$ is the measured flux in the $2-100$ keV energy
band, $F_{0.1-100}$ is the unabsorbed flux in the $0.1-100$ keV energy band
(they are given in units of $10^{-11}$ erg s$^{-1}$ cm$^{-2}$). It is
interesting to check whether the derived emitting-plasma parameters are
consistent with the assumption that the source being studied is an
intermediate polar. Using the normalization obtained, we can estimate the
radius of the accretion column on the white dwarf surface based on its
simplest single-zone model (Warner 1995). Based on this model, we can
determine $n_e$ and the column height $h_s$ by assuming the composition of
the accreting plasma to be solar:

\begin{equation}
     n_{e} = 3.86\cdot 10^{15}~a~m^{-1/2}~R_9^{1/2}~~\rm cm^{-3},
\end{equation}

\begin{equation}
     h_{\rm s} = 1.92\cdot 10^{8}~a^{-1}~m^{3/2}~R_9^{-3/2}~~\rm cm.
\end{equation}

\noindent
where $m$ is the white dwarf mass expressed in solar masses, $a$ is the
local accretion rate in g s$^{-1}$ cm$^{-2}$, $R_9$ is the white dwarf
radius in units of $10^9$ cm. The accretion column radius is then

\begin{equation}
R_{\rm AC} \approx 1.4\cdot10^{7}~d_{100}~a^{-1/2}~m^{-1/4}~R_9^{1/4}~~\rm cm,
\end{equation}

i.e., several percent of the white dwarf radius, in good agreement with the
present views. Here, we use the designation $d_{100} = D/100$ pc.

For a more physical description of the broadband spectrum for
IGR\,J16547$-$1916, we used a grid of accretion column model spectra from
Suleimanov et al. (2005) (see also Revnivtsev et al. 2004) in which the
white dwarf mass is the only parameter. The maximum possible plasma
temperature in the accretion column and, hence, the characteristic energy of
the exponential cutoff in the source energy spectrum depends on the white
dwarf mass. Despite a moderate quality of the hard X-ray spectroscopic data,
we can roughly estimate the mass of the white dwarf in the binary system:
$M_{\rm WD}\simeq0.85\pm0.15 M_{\odot}$. The best-fit parameters are given
below (the notation is the same as that above):

\medskip

\begin{tabular}{c|c|c|c}
$M_{\rm WD}/M_{\odot}$ & $N_{H}$, 10$^{22}$ cm$^{-2}$ & $C_{F}$ & $\chi^2$ \\[1mm]

\hline
$0.85^{+0.15}_{-0.15}$ & $8.4^{+1.4}_{-1.2}$ & $0.961^{+0.009}_{-0.008}$ &  0.83 \\
\end{tabular}

\medskip

\section{CONCLUSIONS}

Here, we presented the results of X-ray and optical observations of the
source IGR\,J16547-1916 from the INTEGRAL survey. We showed the following:

-- The optical spectrum of the source has all the signatures typical of
cataclysmic variables (accreting white dwarfs) -- a blue continuum with a
series of intense Balmer emission lines and a distinct HeII $4686\AA$ line.

-- Our analysis of the optical variability of the source revealed coherent
(as far as this can be judged from our relatively short observations)
brightness oscillations with a period of $549\pm15$ s. We hypothesize that
this period is the spin period of the white dwarf.  Hence it follows that
the source belongs to the class of intermediate polars, i.e., accreting
white dwarfs whose magnetic field is strong enough to destroy the disk flow
near the white dwarf surface but is not strong enough to completely prevent
it formation.

-- The broadband X-ray ($0.6-120$ keV) spectrum of the source is well
described by a model typical of intermediate polars.the model of optically
thin emission from a plasma heated in a shock wave near the white dwarf
surface. The shape of the source spectrum at energies above $10-20$ keV
allows a simple estimate of the white dwarf mass to be made: $M_{\rm
WD}\simeq0.85\pm0.15 M_{\odot}$.

\section{ACKNOWLEDGMENTS}

We wish to thank T. Fatkhullin for help in conducting the observations with
the BTA (SAO) telescope and S. Tsygankov for help in reconstructing the
source spectrum from INTEGRAL data. This work was supported by the Russian
Foundation for Basic Research (project nos. 10-02-00492, 10-02-01442,
10-02-91223, 09-02-97013-p-povolzh'ye-à), the Russian Academy of Sciences
(the ''Origin, Structure, and Evolution of Objects in the Universe'' and
''Active and Stochastic Processes in the Universe'' Program), the Program of
the President of Russia for support of scientific schools (grant no. NSh-
5069.2010.2) and the government contract 14.740.11.0611.  Authors thank to
Shtaerman for the translation of the paper.

\pagebreak

\centerline{REFERENCES}

1. V. L. Afanasiev and A. V. Moiseev, Astron. Lett. 31, 194 (2005).

2. I. F. Bikmaev, R. A. Burenin, M. G. Revnivtsev, et al., Astron. Lett. 34,
653 (2008).

3. I. F. Bikmaev, M. G. Revnivtsev, R. A. Burenin, and
R. A. Sunyaev, Astron. Lett. 32, 588 (2006).

4. A. Bird, A. Bazzano, L. Bazzani, et al., Astrophys.
J. Suppl. Ser. 186, 1 (2010).

5. R. A. Burenin, I. F. Bikmaev,M. G. Revnivtsev, et al., Astron. Lett. 35, 71
(2009).

6. R. A. Burenin, A. V. Meshchepyakov, M. G. Revnivtsev,
et al., Astron. Lett. 34, 367 (2008).

7. H.-J. Grimm, M. Gilfanov, and R. Sunyaev, Astron.
Astrophys. 391, 923 (2002).

8. P. Hertz and J. Grindlay, Astrophys. J. 278, 137
(1984).

9. R. Krivonos, M. Revnivtsev, A. Lutovinov, et al., Astron.
Astrophys. 475, 775 (2003).

10. R. Krivonos, S. Tsygankov, M. Revnivtsev,
et al., Astron. Astrophys. (2010, in press);
http://arxiv.org/abs/1006.4437.

11. N. Masetti, R. Landi, M. Pretorius, et al., Astron.
Astrophys. 470, 331 (2007).

12. N. Masetti, P. Parisi, E. Palazzi, et al.,
Astron. Astrophys. (2010, in press);
http://arxiv.org/abs/1006.4513.

13. J. Patterson, Publ. Astron. Soc. Pacif. 106, 209
(1994).

14. M. G. Revnivtsev, A. Yu. Knyazev, S. Yu. Sazonov,
et al., Astron. Lett. 35, 35 (2009).

15. M. G. Revnivtsev, A. A. Lutovinov, V. F. Suleimanov,
et al., Astron. Lett. 30, 772 (2004).

16. M. Revnivtsev, A. Lutovinov, E. Churazov, et al., Astron.
Astrophys. 491, 209 (2008).

17. S. Sazonov,M. Revnivtsev, R. Burenin, et al., Astron.
Astrophys. 487, 509 (2008).

18. V. Suleimanov, M. Revnivtsev, and H. Ritter, Astron.
Astrophys. 435, 191 (2005).

19. J. Tomsick, S. Chaty, J. Rodriguez, et al., Astrophys.
J. 685, 1143 (2008).

20. J. Tomsick, S. Chaty, J. Rodriguez, et al., Astrophys.
J. 701, 811 (2009).

21. F. Verbunt, W. Bunk, H. Ritter, and E. Pfeffermann,
Astron. Astrophys. 327, 602 (1997).

22. R. E. Williams and D. H. Ferguson, Astrophys. J.
257, 672 (1982).

23. C. Winkler,T. Courvoisier, G. Di Cocco, et al.,A stron.
Astrophys. 411, L1 (2003).

\newpage

\begin{center}
  \begin{figure}
    \centering
    \includegraphics[width=12cm]{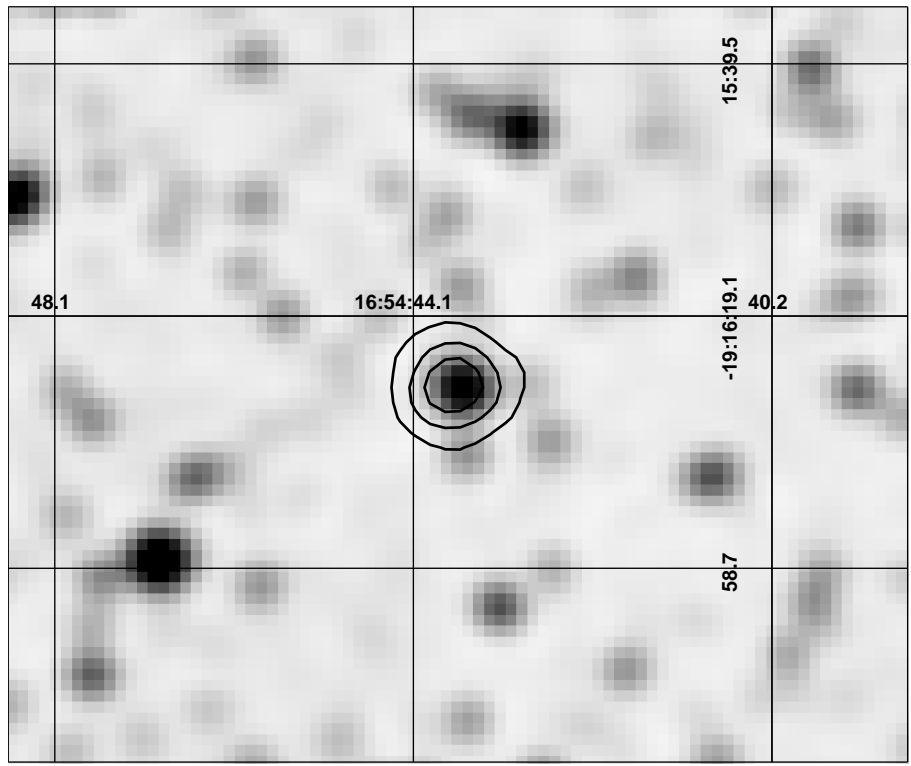}
  \caption{Image of the sky region containing the source IGR\,J16547$-$1916 in the DSS. The contours indicate the intensity levels of the source in X-rays from SWIFT data.}
  \end{figure}\label{optima}
\end{center}

\newpage

\begin{center}
  \begin{figure}
    \smfigure{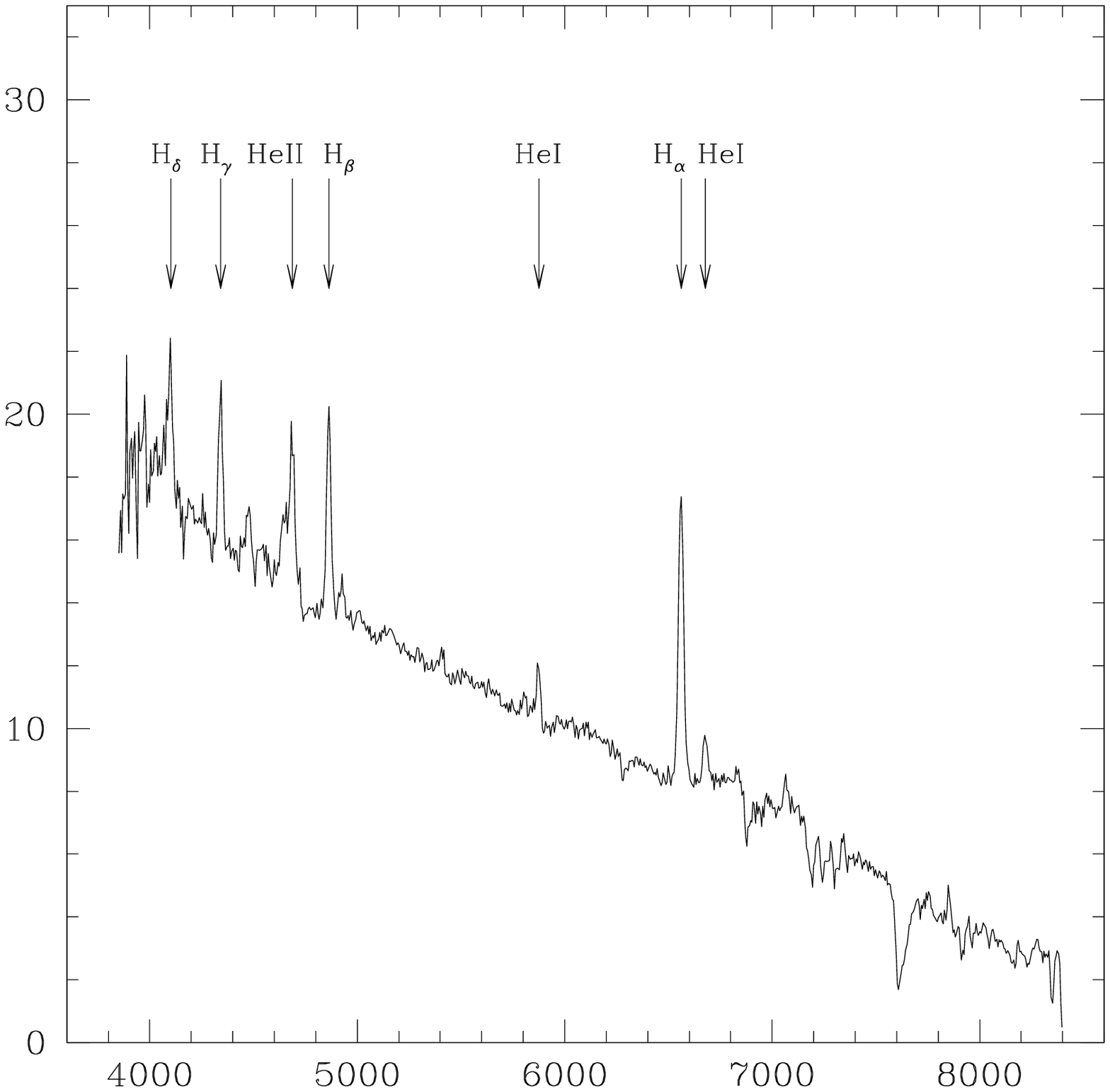}{$\lambda$, $\AA$}{$F_\lambda, \times
      10^{-16}$~erg~s$^{-1}$~cm$^{-2}$~$\AA^{-1}$}
  \caption{Optical spectrum of IGR\,J16547$-$1916 taken with the BTA (SAO) telescope on June 9, 2010. The main lines detected in the spectrum are shown.}
  \end{figure}
\end{center}

\newpage

\begin{center}
\begin{figure}
\includegraphics[width=15cm]{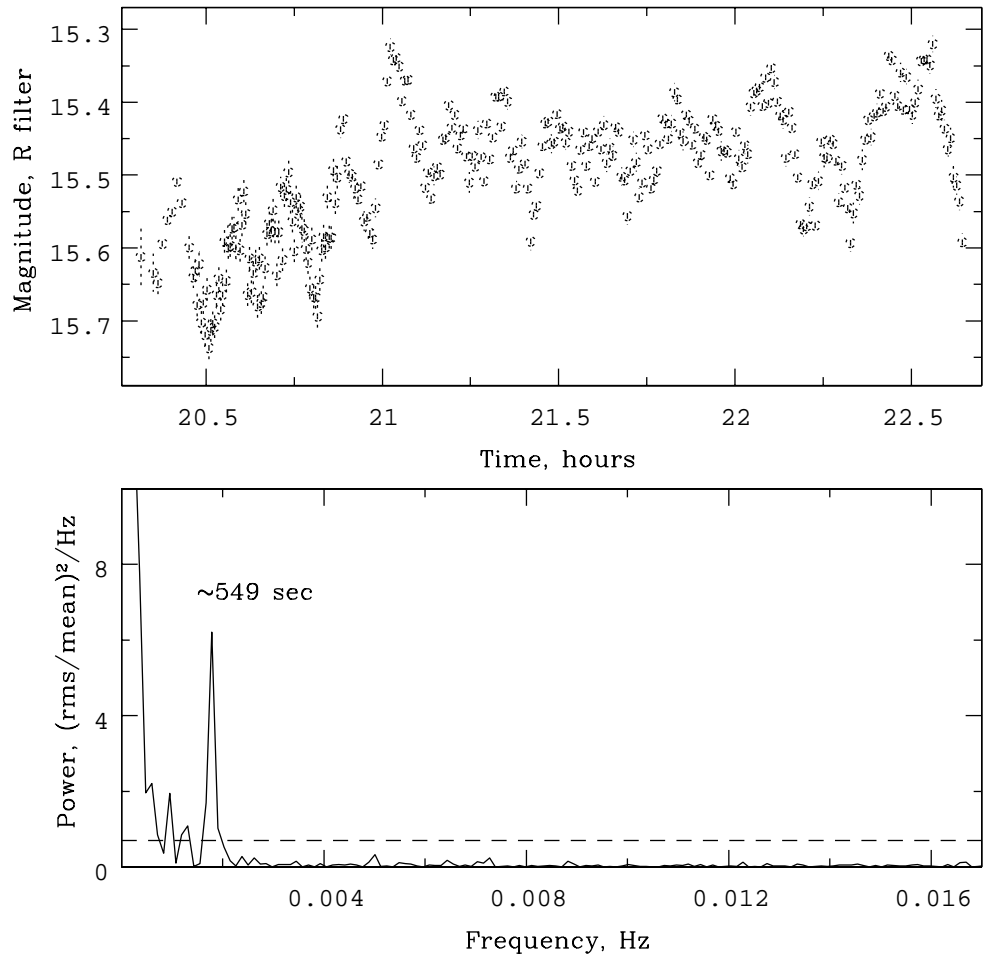}
\caption{(a) Optical light curve of IGR\,J16547$-$1916 obtained with the RTT-150 telescope on June 29, 2010. (b) Lomb-Scargle periodogram of the light curve from panel (a) of the Fig. The dashed line indicates the $5\sigma$ confidence level}
\end{figure}
\end{center}

\newpage

\begin{center}
\begin{figure}
\includegraphics[width=14cm]{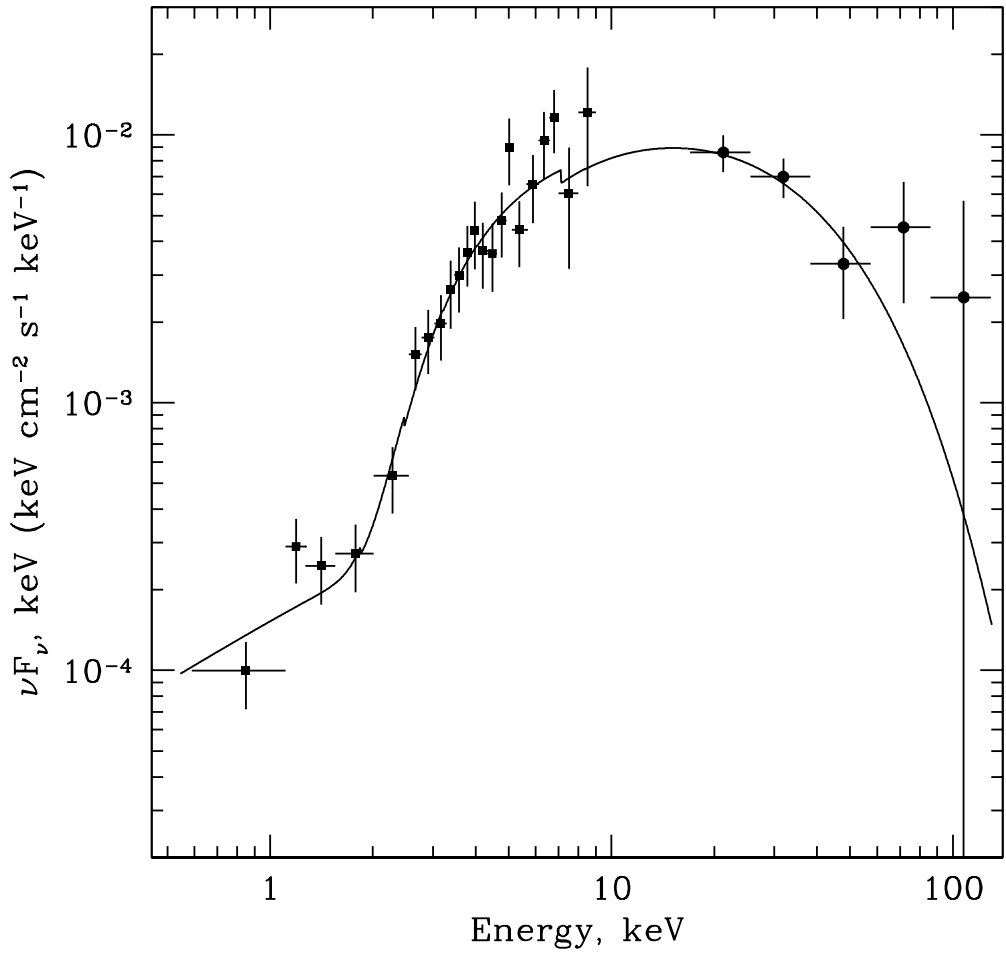}

\caption{IGR\,J16547$-$1916 broad-band energy spectrum, derived from the
data by SWIFT (squares) and INTEGRAL (circles) observatories over the
0.6$-$120 keV spectral range. Solid line shows the best approximation by
partially absorbed bremsstrahlung emission model.}

\end{figure}\label{xrayspec}

\end{center}

\end{document}